\UseRawInputEncoding
\documentclass[AMA,STIX1COL]{WileyNJD-v2}

\articletype{Article Type}%

\received{August 2021}
\revised{---}
\accepted{----}

\begin{document}

\title{Torques and angular momenta of fluid elements in the octonion spaces}

\author[1,2]{Zi-Hua Weng*}

%\author[2,3]{Author Two}

%\author[3]{Author Three}

\authormark{WENG}

\address[1]{\orgdiv{School of Aerospace Engineering}, \orgname{Xiamen University}, \orgaddress{\state{Xiamen 361005}, \country{China}}}

\address[2]{\orgdiv{College of Physical Science and Technology}, \orgname{Xiamen University}, \orgaddress{\state{Xiamen 361005}, \country{China}}}

%\address[3]{\orgdiv{Org Division}, \orgname{Org Name}, \orgaddress{\state{State name}, \country{Country name}}}

\corres{*School of Aerospace Engineering, Xiamen University, Xiamen 361005, China. \email{xmuwzh@xmu.edu.cn}}

%\presentaddress{This is sample for present address text this is sample for present address text}

\abstract[Summary]{
The paper focuses on applying the octonions to explore the influence of the external torque on the angular momentum of fluid elements, revealing the interconnection of the external torque and the vortices of vortex streets. J. C. Maxwell was the first to introduce the quaternions to study the physical properties of electromagnetic fields. The contemporary scholars utilize the quaternions and octonions to investigate the electromagnetic theory, gravitational theory, quantum mechanics, special relativity, general relativity and curved spaces and so forth. The paper adopts the octonions to describe the electromagnetic and gravitational theories, including the octonionic field potential, field strength, linear momentum, angular momentum, torque and force and so on. In case the octonion force is equal to zero, it is able to deduce eight independent equations, including the fluid continuity equation, current continuity equation, and force equilibrium equation and so forth. Especially, one of the eight independent equations will uncover the interrelation of the external torque and angular momentums of fluid elements. One of its inferences is that the direction, magnitude and frequency of the external torque must impact the direction and curl of the angular momentum of fluid elements, altering the frequencies of Karman vortex streets within the fluids. It means that the external torque is interrelated with the velocity circulation, by means of the liquid viscosity. The external torque is able to exert an influence on the direction of downwash flows, improving the lift and drag characteristics generated by the fluids.
}


\jnlcitation{\cname{%
\author{Weng Z.-H.},
%\author{B. Hoskins},
%\author{R. Lee},
%\author{G. Masato}, and
%\author{T. Woollings}
} (\cyear{2021}),
\ctitle{Torques and angular momenta of fluid elements in the octonion spaces},
\cjournal{Math Meth Appl Sci.},
\cvol{2021;00:1--6}.}

\maketitle

%\footnotetext{\textbf{Abbreviations:} ANA, anti-nuclear antibodies; APC, antigen-presenting cells; IRF, interferon regulatory factor}

\section{\label{sec:level1}Introduction}

Why is the Karman vortex street staggered? Why are the vorticity directions of vortices different in the vortex streets? Can the frequency of staggered vortices be changed? Which factors may exert an influence on the divergence of angular momentum of the fluid element? For a long time, these interesting questions have attracted and puzzled scholars. It was not until the emergence of electromagnetic and gravitational theories, described with the octonions, that these questions were answered to some extent. According to this field theory, the periodically varying torque is able to induce the angular momentum of vortexes to change periodically, under the pressure difference produced by the low-velocity fluids on both sides of the vortex street. Changing the frequency of external torque is capable of altering the frequency of vorticity. The divergence of the angular momentum of the vortex can be transformed by modifying the magnitude of external torque. These physical properties will help to improve the lift and drag characteristics generated by the fluids.

In the fluid mechanics, the divergence of angular momentum of the fluid element plays an important role. a) Typhoon track. In the weather forecast, it is of great theoretical significance and practical value to predict the possible course of typhoon. After intensive research, scientists have concluded that the variation of flux divergence of the angular momentum, which is relevant to the air flow in the atmosphere, has a direct influence on the path of typhoon. b)Vortex street. In the experiment of Karman vortex street \cite{song}, besides the divergence of angular momentum of the fluid elements, the torque, that the fluid is subjected to, has an important influence on the direction of motion of the fluid, flowing through the surface of a cylinder. c) Cloud pattern. When clouds flow through mountains (or skyscrapers), the vortex street patterns also appear in clouds. Mountains (or skyscrapers) are able to exert torque on clouds, transforming the divergence and curl of angular momentum of clouds, impacting the discrete vortex of clouds.

However, the existing fluid mechanics is unable to investigate effectively the interrelationship of the fluid movement with the flux divergence of the angular momentum of fluid elements. Either it is incapable of revealing the influence of the electromagnetic strength and gravitational strength on the motion characteristics of the fluid (water, cloud), magneto-fluid \cite{tanisli1}, and plasma and so forth. Obviously, this theoretical limitation confines the range of application of existing fluid mechanics.

In sharp contrast, the electromagnetic and gravitational theory, described with the octonions (octonion field theory, for short), can explicate the formation of Karman vortex street. This field theory uncovers the contribution of external torque, electromagnetic strength, and gravitational strength and others on the flux divergence of the angular momentum of fluid elements. And it extends the scope of application of fluid mechanics.

J. C. Maxwell was the first to apply the quaternions to describe the electromagnetic theory. Nowadays, the scholars utilize the classical and non-classical octonions to research the electromagnetic theory \cite{mironov,gogberashvili1}, gravitational theory, dark matter \cite{furui1,furui2}, quantum mechanics \cite{deleo1,deleo2,gogberashvili2}, quantum Hall effect \cite{bernevig}, special relativity \cite{manogue}, general relativity \cite{bossard}, differential geometry \cite{beggs}, and strong-unclear theory \cite{chanyal2,furey}, and weak-unclear theory \cite{majid,farrill} and so forth.

The electromagnetic and gravitational theory can be described by octonions \cite{weng1} , especially the octonion force (in Section 2). In case the octonion force is equals to zero, it is able to achieve eight equations independent of each other. One of these equations claims that the external torque (in Section 4) will make a contribution on the angular momentum of fluid elements, including the direction, divergence, and frequency and so forth.

1) Torque and velocity circulation. In the experiment of velocity circulation, making use of the liquid viscosity, the rotating cylinder can transfer the torque to the fluid surrounding it, impacting the direction of fluid flow. Apparently, there is a direct interrelation between the external torque and velocity circulation of fluid. In other words, the external torque is directly related to the lift generated by the fluid.

2) Divergence and curl. In the preceding experiment of velocity circulation, the external torque may impact the angular momentum of fluids. The torque that the rotating cylinder transfers to the fluids can alter the angular momentum of fluid elements, including the magnitude, curl, and divergence. By the variation of external torque, one can modify the curl and divergence of the angular momentum of fluid elements, shifting the direction of fluid flow. Obviously, there is a relation between the external torque and the divergence of angular momentum of fluids.

3) Frequency and vortex street. According the torque continuity equation (in Section 4), if the external torque is a periodic function, the induced angular momentum and vortex street must be periodic functions also. Similarly, the relevant velocity circulation and lift will be shifted periodically too. Further, modifying the periodic frequency of torque may alter that of angular momentum (or velocity circulation, or lift).

What the paper studies is the classical octonion, which was invented by T. Graves and A. Cayley independently. In order to study gravi-electromagnetism \cite{chanyal3,tanisli2}, dark matter \cite{chanyal4}, dyons \cite{kansu,demir} and others, some scholars developed the non-classical octonions, including the hyperbolic-octonions, split-octonions, pseudo-octonions, Cartan's octonions, and others. The classical octonions and the non-classical octonions complement each other to a certain extent. By means of the electromagnetic and gravitational theories described with the octonions, the paper discusses the causes and influencing factors of Karman vortex street in the fluids, from the interrelationships of external torque and divergence of angular momentum of fluid elements.

\begin{table}[h]
\caption{Some subspaces and physical quantities in the electromagnetic and gravitational fields described with the octonions.}
%\label{sphericcase}
\center
\begin{tabular}{@{}ll@{}}
\hline\hline
subspace                                                         &  physical quantity                                                              \\
\hline
quaternion space, $\mathbb{H}_g (\emph{\textbf{i}}_j)$           &  $\mathbb{R}_g$ , $\mathbb{V}_g$ , $\mathbb{A}_g$ ,
                                                                    $\mathbb{X}_g$ , $\mathbb{F}_g$ ,
                                                                    $\mathbb{S}_g$ , $\mathbb{P}_g$ ,
                                                                    $\mathbb{L}_g$ , $\mathbb{W}_g$ , $\mathbb{N}_g$                               \\
second subspace, $\mathbb{H}_{em} (\emph{\textbf{I}}_j)$         &  $\mathbb{R}_{em}$ , $\mathbb{V}_{em}$ , $\mathbb{A}_{em}$ ,
                                                                    $\mathbb{X}_{em}$ , $\mathbb{F}_{em}$ ,
                                                                    $\mathbb{S}_{em}$ , $\mathbb{P}_{em}$ ,
                                                                    $\mathbb{L}_{em}$ , $\mathbb{W}_{em}$ , $\mathbb{N}_{em}$                      \\
\hline\hline
\end{tabular}
\end{table}

\section{Octonion force}

The octonion space $\mathbb{O}$ can be separated into a few subspaces independent of each other, including $\mathbb{H}_g$ and $\mathbb{H}_{em}$ . The subspace $\mathbb{H}_g$ is one quaternion space, which can be applied to depict the physical properties of gravitational fields. Meanwhile, the second subspace $\mathbb{H}_{em}$ may be utilized to describe the physical properties of electromagnetic fields (Table 1).

In the octonion field theory, the gravitational potential is $\mathbb{A}_g$ , in the subspace $\mathbb{H}_g$ . Meanwhile the electromagnetic potential is $\mathbb{A}_{em}$ , in the second subspace $\mathbb{H}_{em}$ . Therefore the octonion field potential is, $\mathbb{A} = \mathbb{A}_g + k_{eg} \mathbb{A}_{em}$ . Herein $k_{eg}$ is the coefficient to ensure the dimensional homogeneity.

From the octonion field potential $\mathbb{A}$ and quaternion operator $\lozenge$ , it is able to define the octonion field strength, $\mathbb{F} = \mathbb{F}_g + k_{eg} \mathbb{F}_{em}$. The octonion field strength $\mathbb{F}$ consists of the gravitational strength $\mathbb{F}_g$ , in the subspace $\mathbb{H}_g$, and the electromagnetic strength $\mathbb{F}_{em}$ , in the second subspace $\mathbb{H}_{em}$ . The gravitational strength $\mathbb{F}_g$ comprises the gravitational acceleration $\textbf{g}$ and gravitational precessional-angular-velocity $\textbf{b}$ \cite{weng2} . And the electromagnetic strength $\mathbb{F}_{em}$ includes the magnetic flux density $\textbf{B}$ and electric field intensity $\textbf{E}$ . Here  $i$ is the imaginary unit. $\emph{\textbf{i}}_j$ is the basis vector in the subspace $\mathbb{H}_g$ , while $\emph{\textbf{I}}_j$ is that in the second subspace $\mathbb{H}_{em}$ . $\lozenge = i \partial_0 + \nabla$ . $\nabla = \textbf{i}_k \partial_k$. $\partial_j = \partial / \partial r_j $ . $r_j$ is the coordinate value, in the subspace $\mathbb{H}_g$ . $r_0 = v_0 t$ . $t$ is the time. $\emph{\textbf{i}}_0 = 1$ , $\emph{\textbf{i}}_k^2 = -1$ , $\emph{\textbf{I}}_j^2 = -1$ . $\emph{\textbf{I}}_j = \emph{\textbf{i}}_j \circ \emph{\textbf{I}}_0$ . $j = 0, 1, 2, 3$. $k = 1, 2, 3$.

Similarly, from the octonion field strength $\mathbb{F}$ and quaternion operator $\lozenge$, one can define the octonion field source, $ \mu \mathbb{S} = - ( i \mathbb{F} / v_0 + \lozenge )^\ast \circ \mathbb{F}$ . Further it can be separated as follows \cite{weng3},
\begin{eqnarray}
\mu \mathbb{S} = \mu_g \mathbb{S}_g + k_{eg} \mu_e \mathbb{S}_{em} -  i \mathbb{F}^\ast \circ \mathbb{F} / v_0 ~ ,
\end{eqnarray}
where the octonion field source $\mathbb{S}$ comprises the gravitational source $\mathbb{S}_g$ , in the subspace $\mathbb{H}_g$ , and the electromagnetic source $\mathbb{S}_{em}$ , in the second subspace $\mathbb{H}_{em}$. Further, the definition of electromagnetic source, $\mathbb{S}_{em}$ , is able to deduce the electromagnetic field equations, which is identical of the Maxwell's equations. The definition of gravitational source, $\mathbb{S}_g$ , is capable of inferring the gravitational field equations, which covers the Newton's law of universal gravitation. $\mu$ is a coefficient, $v_0$ is the speed of light. $\mu_g$ and $\mu_e$ are the gravitational constant and electromagnetic constant, respectively. $\ast$ is the octonion conjugate. The symbol $\circ$ denotes the octonion multiplication (Table 2).

\begin{table}[h]
\caption{Some physical quantities and definitions of the gravitational and electromagnetic fields.}
\center
\begin{tabular}{@{}ll@{}}
\hline\hline
physical quantity                      &   definition                                                                                                                \\
\hline
quaternion operator                    &   $\lozenge = i \partial_0 + \nabla$                                                                                        \\
octonion radius vector                 &   $\mathbb{R} = \mathbb{R}_g + k_{eg} \mathbb{R}_{em}$                                                                      \\
octonion integrating function          &   $\mathbb{X} = \mathbb{X}_g + k_{eg} \mathbb{X}_{em}$                                                                      \\
octonion field potential               &   $\mathbb{A} = i \lozenge^\times \circ \mathbb{X}$                                                                         \\
octonion field strength                &   $\mathbb{F} = \lozenge \circ \mathbb{A}$                                                                                  \\
octonion field source                  &   $\mu \mathbb{S} = - ( i \mathbb{F} / v_0 + \lozenge )^\ast \circ \mathbb{F}$                                              \\
octonion linear momentum               &   $\mathbb{P} = \mu \mathbb{S} / \mu_g$                                                                                     \\
octonion angular momentum              &   $\mathbb{L} = ( \mathbb{R} + k_{rx} \mathbb{X} )^\times \circ \mathbb{P} $                                                \\
octonion torque                        &   $\mathbb{W} = - v_0 ( i \mathbb{F} / v_0 + \lozenge ) \circ \{ ( i \mathbb{V}^\times / v_0 ) \circ \mathbb{W} \} $        \\
octonion force                         &   $\mathbb{N} = - ( i \mathbb{F} / v_0 + \lozenge ) \circ \{ ( i \mathbb{V}^\times / v_0 ) \circ \mathbb{W} \} $            \\
\hline\hline
\end{tabular}
\end{table}

From the octonion field source $\mathbb{S}$ , one can define the octonion linear momentum, $\mathbb{P} = \mu \mathbb{S} / \mu_g$ . The octonion linear momentum consists of the component $\mathbb{P}_g$ , in the subspace $\mathbb{H}_g$ , and the component $\mathbb{P}_{em}$ , in the second subspace $\mathbb{H}_{em}$, that is, $\mathbb{P} = \mathbb{P}_g + k_{eg} \mathbb{P}_{em}$.

The octonion radius vector $\mathbb{R}$ comprises the component $\mathbb{R}_g$ in the subspace $\mathbb{H}_g$ , and the component $\mathbb{R}_{em}$ in the second subspace $\mathbb{H}_{em}$ . The octonion integrating function, $\mathbb{X}$ , of field potential includes the component $\mathbb{X}_g$ in the subspace $\mathbb{H}_g$, and the component $\mathbb{X}_{em}$ in the second subspace $\mathbb{H}_{em}$ . Moreover, the octonion field potential $\mathbb{A}$ can be defined from the octonion integrating function $\mathbb{X}$ and quaternion operator $\lozenge$ , that is, $\mathbb{A} = i \lozenge^\times \circ \mathbb{X}$ , with $\times$ being the complex conjugate.

Making use of the octonion linear momentum $\mathbb{P}$ and radius vector $\mathbb{R}$ , it is capable of the octonion angular momentum, $\mathbb{L} = (\mathbb{R} + k_{rx} \mathbb{X} )^\times \circ \mathbb{P}$ . And the latter can be separated into,
\begin{eqnarray}
\mathbb{L} = \mathbb{L}_g + k_{eg} \mathbb{L}_{em}  ~ ,
\end{eqnarray}
where $\mathbb{L}_g$ is the component in the subspace $\mathbb{H}_g$ , while $\mathbb{L}_{em}$ is the component in the second subspace $\mathbb{H}_{em}$ . $\mathbb{L}_g = L_{10} + \emph{i} \textbf{L}_1^\textrm{i} + \textbf{L}_1$. $\textbf{L}_1$ includes the angular momentum density, and $\textbf{L}_1^i$ is called as the mass moment temporarily. $\mathbb{L}_{em} = \textbf{L}_{20} + \emph{i} \textbf{L}_2^\textrm{i} + \textbf{L}_2$. $\textbf{L}_2^\textrm{i}$ covers the electric momentum. $\textbf{L}_2 $ includes the magnetic momentum. $\textbf{L}_1 = \Sigma L_{1k} \emph{\textbf{i}}_k$, $\textbf{L}_1^\textrm{i} = \Sigma L_{1k}^\textrm{i} \emph{\textbf{i}}_k$. $\textbf{L}_{20} = L_{20} \emph{\textbf{I}}_0$, $\textbf{L}_2 = \Sigma L_{2k} \emph{\textbf{I}}_k$, $\textbf{L}_2^\textrm{i} = \Sigma L_{2k}^\textrm{i} \emph{\textbf{I}}_k$. $L_{1j}$ , $L_{1k}^\textrm{i}$ , $L_{2j}$ , and $L_{2k}^\textrm{i}$ are all real. $k_{rx}$ is a coefficient, to ensure the dimensional homogeneity.

From the octonion angular momentum $\mathbb{L}$ and quaternion operator $\lozenge$ , it is able to define the octonion torque as follows,
\begin{eqnarray}
\mathbb{W} = - v_0 ( i \mathbb{F} / v_0 + \lozenge ) \circ \{ ( i \mathbb{V}^\times / v_0 ) \circ \mathbb{L} \} ~ ,
\end{eqnarray}
where $\mathbb{W}_g$ and $\mathbb{W}_{em}$ are the components of octonion torque, in the subspace $\mathbb{H}_g$ and second subspace $\mathbb{H}_{em}$ , respectively, that is, $\mathbb{W} = \mathbb{W}_g + k_{eg} \mathbb{W}_{em}$. Further, $\mathbb{W}_g = i W_{10}^\textrm{i} + W_{10} + i \textbf{W}_1^\textrm{i} + \textbf{W}_1$ . $W_{10}^\textrm{i}$ is the energy. $\textbf{W}_1^\textrm{i}$ is the torque, including the gyroscopic torque, $i \nabla ( \textbf{v} \cdot \textbf{L}_1 )$ . $W_{10}$ is the divergence of angular momentum, and $\textbf{W}_1$ is the curl of angular momentum. Similarly, $\mathbb{W}_{em} = i \textbf{W}_{20}^\textrm{i} + \textbf{W}_{20} + i \textbf{W}_2^\textrm{i} + \textbf{W}_2$ . $\textbf{W}_{20}^\textrm{i}$ and $\textbf{W}_2^\textrm{i}$ are called as the second-energy and second-torque temporarily and respectively. $\textbf{W}_{20}$ is the divergence of magnetic moment. $\textbf{W}_2$ covers the curl of magnetic moment and derivative of electric moment. $\textbf{v} = \Sigma \textbf{i}_k v_k$ . $\textbf{W}_1 = \Sigma W_{1k} \emph{\textbf{i}}_k$, $\textbf{W}_1^\textrm{i} = \Sigma W_{1k}^\textrm{i} \emph{\textbf{i}}_k$. $\textbf{W}_{20} = W_{20} \emph{\textbf{I}}_0$, $\textbf{W}_2 = \Sigma W_{2k} \emph{\textbf{I}}_k$, $\textbf{W}_2^\textrm{i} = \Sigma W_{2k}^\textrm{i} \emph{\textbf{I}}_k$. $W_{1j}$ , $W_{1j}^\textrm{i}$ , $W_{2j}$ , and $W_{2j}^\textrm{i}$ are all real.

\begin{table}[h]
\caption{Comparison of some physical quantities between the gravitational field and electromagnetic field.}
%\label{tab:3}       % Give a unique label��
\center
\begin{tabular}{@{}lll@{}}
\hline\hline
physical quantity                  &   gravitational field                                          &   electromagnetic field                                         \\
\hline
field strength                     &   gravitational acceleration, $\textbf{g}$                     &   electric field intensity, $\textbf{E}$                        \\
                                   &   gravitational precessional-angular-velocity, $\textbf{b}$    &   magnetic induction intensity, $\textbf{B}$                    \\
field source                       &   mass, $s_0$                                                  &   electric charge, $\textbf{S}_0$                               \\
                                   &   linear momentum, $\textbf{s}$                                &   electric current, $\textbf{S}$                                \\
angular momentum                   &   dot product, $L_{10}$                                        &   dot product, $\textbf{L}_{20}$                                \\
                                   &   angular momentum, $\textbf{L}_1$                             &   magnetic moment, $\textbf{L}_2$                               \\
                                   &   mass moment, $\textbf{L}_1^\textrm{i}$                       &   electric moment, $\textbf{L}_2^\textrm{i}$                    \\
torque                             &   divergence of angular momentum, $W_{10}$                     &   divergence of magnetic moment, $\textbf{W}_{20}$              \\
                                   &   energy, $W_{10}^\textrm{i}$                                  &   second-energy, $\textbf{W}_{20}^\textrm{i}$                   \\
                                   &   curl of angular momentum, $\textbf{W}_1$                     &   curl of magnetic moment, $\textbf{W}_2$                       \\
                                   &   torque, $\textbf{W}_1^\textrm{i}$                            &   second-torque, $\textbf{W}_2^\textrm{i}$                      \\
force                              &   power, $N_{10}$                                              &   second-power, $\textbf{N}_{20}$                               \\
                                   &   torque divergence, $N_{10}^\textrm{i}$                       &   second-torque divergence, $\textbf{N}_{20}^\textrm{i}$        \\
                                   &   torque derivative, $\textbf{N}_1$                            &   second-torque derivative, $\textbf{N}_2$                      \\
                                   &   force, $\textbf{N}_1^\textrm{i}$                             &   second-force, $\textbf{N}_2^\textrm{i}$                       \\
\hline\hline
\end{tabular}
\end{table}

By means of the octonion torque $\mathbb{W}$ and quaternion operator $\lozenge$ , it is capable of defining the octonion force as follows,
\begin{eqnarray}
\mathbb{N} = - ( i \mathbb{F} / v_0 + \lozenge ) \circ \{ ( i \mathbb{V}^\times / v_0 ) \circ \mathbb{W} \} ~ ,
\end{eqnarray}
where $\mathbb{N}_g$ and $\mathbb{N}_{em}$ are the components of octonion force, in the two subspaces $\mathbb{H}_g$ and $\mathbb{H}_{em}$ , respectively, that is, $\mathbb{N} = \mathbb{N}_g + k_{eg} \mathbb{N}_{em}$. Further, $\mathbb{N}_g = i N_{10}^\textrm{i} + N_{10} + i \textbf{N}_1^\textrm{i} + \textbf{N}_1$ . $N_{10}^\textrm{i}$ is the divergence of torque, and relevant to the torque continuity equation. $\textbf{N}_1^\textrm{i}$ is the force, including the Magnus force, $ \nabla ( v_0 \partial_0 L_{10} )$ . And it is clearly related to the force equilibrium equation. $N_{10}$ is the power, and relation to the fluid continuity equation. $\textbf{N}_1$ is the derivative of torque, and involved with the precession equilibrium equation. Similarly, $\mathbb{N}_{em} = i \textbf{N}_{20}^\textrm{i} + \textbf{N}_{20} + i \textbf{N}_2^\textrm{i} + \textbf{N}_2$ . $\textbf{N}_{20}$ is called as the second-power temporarily, which is connected with the current continuity equation. $\textbf{N}_2^\textrm{i}$ is called as the second-force temporarily. $\textbf{N}_1 = \Sigma N_{1k} \emph{\textbf{i}}_k$, $\textbf{N}_1^\textrm{i} = \Sigma N_{1k}^\textrm{i} \emph{\textbf{i}}_k$. $\textbf{N}_{20} = N_{20} \emph{\textbf{I}}_0$, $\textbf{N}_2 = \Sigma L_{2k} \emph{\textbf{I}}_k$, $\textbf{N}_2^\textrm{i} = \Sigma N_{2k}^\textrm{i} \emph{\textbf{I}}_k$. $N_{1j}$, $N_{1j}^\textrm{i}$ , $N_{2j}$ , and $N_{2j}^\textrm{i}$ are all real (Table 3).

When the octonion force is equal to zero, that is, $\mathbb{N} = 0$, it is able to achieve eight equations independent to each other, including four continuity equations and four equilibrium equations.

\section{Equilibrium and continuity equations}

In the octonion space, in case the octonion force $\mathbb{N}$ equals to zero, it is capable of inferring eight independent equations. Four of eight equations possess distinct physical meaning, including the fluid (or mass, or linear momentum) continuity equation, current continuity equation, force equilibrium equation, and precession equilibrium equation (Table 4). The remaining four equations need to be further explored in terms of their respective physical meanings. According to the octonion field theory, the electromagnetic strength ($\textbf{E}$, $\textbf{B}$) and gravitational strength ($\textbf{g}$, $\textbf{b}$) may make a contribution to the eight equations (Table 5) to a certain extent, in the electromagnetic and gravitational fields.

1) Fluid continuity equation. In the octonion space $\mathbb{O}$ , for the electromagnetic and gravitational fields, the fluid continuity equation, $N_{10} = 0$, can be written as,
\begin{eqnarray}
0 = && ( \textbf{g} \cdot \textbf{W}_1 / v_0 + \textbf{b} \cdot \textbf{W}_1^\textrm{i} ) / v_0 + ( \partial_0 W_{10}^\textrm{i} -  \nabla \cdot \textbf{W}_1 )
\nonumber
\\
&&
+ k_{eg}^2 ( \textbf{E} \cdot \textbf{W}_2/ v_0 + \textbf{B} \cdot \textbf{W}_2^\textrm{i} ) / v_0  ~  .
\end{eqnarray}

When there is neither electromagnetic strength nor gravitational strength, the above can be reduced into the traditional fluid continuity equation in the classical fluid mechanics. However, no one has ever verified/invalidated the contribution of electromagnetic and/or gravitational strength to the fluid continuity equation, in the classical fluid mechanics up to now.

2) Current continuity equation. In the octonion space for the electromagnetic and gravitational fields, the current continuity equation, $\textbf{N}_{20} = 0$, will be expanded into,
\begin{eqnarray}
0 = && ( \textbf{g} \cdot \textbf{W}_2 / v_0 + \textbf{b} \cdot \textbf{W}_2^\textrm{i} ) / v_0
+ ( \partial_0 \textbf{W}_{20}^\textrm{i} - \nabla \cdot \textbf{W}_2 )
\nonumber
\\
&&
+ ( \textbf{E} \cdot \textbf{W}_1 / v_0 + \textbf{B} \cdot \textbf{W}_1^\textrm{i} ) / v_0 ~.
\end{eqnarray}

In case there is neither electromagnetic strength nor gravitational strength, the above may be simplified into the traditional current continuity equation in the classical electromagnetic theory. However, no one has ever validated the contribution of electromagnetic and/or gravitational strength to the current continuity equation, in the classical electromagnetic theory so far.

Moreover, in the octonion space, the establishment of current continuity equation states that the existing sciences and technologies are able to measure some physical quantities and relevant equations in the second subspace $\mathbb{H}_{em}$ . Further, it is possible to discover more physical quantities and relevant equations in the second subspace $\mathbb{H}_{em}$ .

3) Force equilibrium equation. In the octonion space for the electromagnetic and gravitational fields, the force equilibrium equation, $\textbf{N}_1^\textrm{i} = 0$, may be rewritten as,
\begin{eqnarray}
0 = && ( W_{10}^\textrm{i} \textbf{g} / v_0 + \textbf{g} \times \textbf{W}_1^\textrm{i} / v_0
- W_{10} \textbf{b} - \textbf{b} \times \textbf{W}_1 ) / v_0
\nonumber
\\
&&
- ( \partial_0 \textbf{W}_1 + \nabla W_{10}^\textrm{i} + \nabla \times \textbf{W}_1^\textrm{i} )
\nonumber
\\
&&
+ k_{eg}^2 ( \textbf{E} \circ \textbf{W}_{20}^\textrm{i} / v_0 + \textbf{E} \times \textbf{W}_2^\textrm{i} / v_0
- \textbf{B} \circ \textbf{W}_{20} - \textbf{B} \times \textbf{W}_2 ) / v_0 ~.
\end{eqnarray}

The above can be degenerated into the traditional force equilibrium equation in the classical field theory. It is found that the electromagnetic strength ($\textbf{E}$, $\textbf{B}$) and gravitational strength ($\textbf{g}$, $\textbf{b}$) both make significant contributions to the force equilibrium equation, according to the octonion field theory or the classical field theory.

4) Precession equilibrium equation. In the octonion space for the electromagnetic and gravitational fields, the precession equilibrium equation, $\textbf{N}_1 = 0$, will be extended into,
\begin{eqnarray}
0 = &&  ( W_{10} \textbf{g} / v_0 + \textbf{g} \times \textbf{W}_1 / v_0
+ W_{10}^\textrm{i} \textbf{b} + \textbf{b} \times \textbf{W}_1^\textrm{i} ) / v_0
\nonumber
\\
&&
+ ( \partial_0 \textbf{W}_1^\textrm{i} - \nabla W_{10} - \nabla \times \textbf{W}_1 )
\nonumber
\\
&&
+ k_{eg}^2 ( \textbf{E} \circ \textbf{W}_{20} / v_0 + \textbf{E} \times \textbf{W}_2/ v_0
+ \textbf{B} \circ \textbf{W}_{20}^\textrm{i} + \textbf{B} \times \textbf{W}_2^\textrm{i} ) / v_0  ~.
\end{eqnarray}

The above is able to deduce some familiar inferences relevant to the precessional motions, including the angular velocity of Larmor precession and so forth. And it points out the contributions of electromagnetic strength ($\textbf{E}$, $\textbf{B}$) and gravitational strength ($\textbf{g}$, $\textbf{b}$) on the precession equilibrium equation.

Nowadays, one of remaining four equations has been verified to be relevant to the external torque and the divergence of angular momentum of fluids. It is what we need to study in the paper.

\begin{table}[h]
\caption{In the octonion field theory, there are eight equilibrium and continuity equations, from $\mathbb{N} = 0$.}
\center
\begin{tabular}{@{}llll@{}}
\hline\hline
No.         &   formula                            &    equilibrium/continuity equation          &  space              \\
\hline
1           &   $\textbf{N}_1^\textrm{i} = 0$      &    force equilibrium equation               &  $\mathbb{H}_g$     \\
2           &   $N_{10} = 0$                       &    fluid continuity equation                &  $\mathbb{H}_g$     \\
3           &   $\textbf{N}_1 = 0$                 &    precession equilibrium equation          &  $\mathbb{H}_g$     \\
4           &   $N_{10}^\textrm{i} = 0$            &    torque continuity equation               &  $\mathbb{H}_g$     \\
5           &   $\textbf{N}_{20} = 0$              &    current continuity equation              &  $\mathbb{H}_{em}$  \\
6           &   $\textbf{N}_2^\textrm{i} = 0$      &    second-force equilibrium equation        &  $\mathbb{H}_{em}$  \\
7           &   $\textbf{N}_2 = 0$                 &    second-precession equilibrium equation   &  $\mathbb{H}_{em}$  \\
8           &   $\textbf{N}_{20}^\textrm{i} = 0$   &    second-torque continuity equation        &  $\mathbb{H}_{em}$  \\
\hline\hline
\end{tabular}
\label{ta1}
\end{table}

\section{Torque continuity equation}

In the octonion space for the electromagnetic and gravitational fields, when the octonion force $\mathbb{N}$ is equals to zero, it is able to achieve eight equilibrium and continuity equations. One of them is, $N_{10}^\textrm{i} = 0$ , which is called as the `torque continuity equation' temporarily. Further it can be written as (see Ref.[18]),
\begin{eqnarray}
0 = && ( \textbf{g} \cdot \textbf{W}_1^\textrm{i} / v_0 - \textbf{b} \cdot \textbf{W}_1 ) / v_0 - ( \partial_0 W_{10} + \nabla \cdot \textbf{W}_1^\textrm{i} )
\nonumber
\\
&&
+ k_{eg}^2 ( \textbf{E} \cdot \textbf{W}_2^\textrm{i} / v_0 - \textbf{B} \cdot \textbf{W}_2 ) / v_0  ~.
\end{eqnarray}

The above can deduce a few special inferences as follows, under certain circumstances.

\subsection{Zero field}

When there is neither electromagnetic strength nor gravitational strength, the torque continuity equation will be reduced into,
\begin{eqnarray}
- \partial ( \nabla \cdot \textbf{L}_1 ) / \partial t + \nabla \cdot \textbf{W}_1^\textrm{i} = 0 ~.
\end{eqnarray}

The above states that the external torque $\textbf{W}_1^\textrm{i}$ is interrelated with the divergence of angular momentum, $\nabla \cdot \textbf{L}_1$ , of fluids, rather than independent of each other. To a certain extent, the torque continuity equation is similar to the fluid continuity equation. By comparison, the divergence of angular momentum and external torque, in the torque continuity equation, correspond to the mass and linear momentum, respectively, in the fluid continuity equation.

It is similar to the variation case of linear momentum, the torque varies in many ways, including the unidirectional change and reciprocating change. This leads to a variety of transformation relevant to the divergences of angular momentum.

Further, the above can be rewritten as,
\begin{eqnarray}
\nabla \cdot \textbf{f}_a = 0 ~,
\end{eqnarray}
where the vector $ \textbf{f}_a =  - \partial \textbf{L}_1 / \partial t + \textbf{W}_1^\textrm{i}$ . There may be some solutions for $\nabla \cdot \textbf{f}_a = 0$, including $\textbf{f}_a = 0$.

According to the torque continuity equation, in case the external torque is a sinusoidal function with the phase value $\omega t$ , the angular momentum must be a cosine function with the phase value $\omega t$ , when $\textbf{f}_a \neq 0$ . The frequencies of the two functions are the same, that is,
\begin{eqnarray}
L_1 = L_{10} cos ( \omega t + \phi ) ~,~~  W_1^i = W_{10}^i sin ( \omega t) ~,
\end{eqnarray}
where $\textbf{L}_1 = \textbf{u} L_1$ , $\textbf{W}_1^i = \textbf{u} W_1^i $ . $\textbf{u}$ is an unit vector in the quaternion space $\mathbb{H}_g$ . $L_{10}$ and $W_{10}^i$ are all real. $\omega$ is the angular velocity. $\phi$ is one phase angle. Obviously, the phase difference between the torque and angular momentum may reach to $\pi / 2$ .

\begin{table}[h]
\caption{In the octonion field theory, the electromagnetic strength and gravitational strength make a contribution to each of eight equilibrium/continuity equations.}
\center
\begin{tabular}{@{}lll@{}}
\hline\hline
  equilibrium/continuity equation          &   octonion theory               &   classical theory                   \\
\hline
  fluid continuity equation                &   Yes                           &   No                                 \\
  torque continuity equation               &   Yes                           &   --                                 \\
  precession equilibrium equation          &   Yes                           &   --                                 \\
  force equilibrium equation               &   Yes                           &   Yes                                \\
  current continuity equation              &   Yes                           &   No                                 \\
  second-torque continuity equation        &   Yes                           &   --                                 \\
  second-precession equilibrium equation   &   Yes                           &   --                                 \\
  second-force equilibrium equation        &   Yes                           &   --                                 \\
\hline\hline
\end{tabular}\label{ta1}
\end{table}

\subsection{Vortex street}

The torque continuity equation can be utilized to explain the cause of Karman vortex street. In the vortex street experiment of viscous-dominated fluids, the torque, produced by the fluid (Reynolds number, $Re < 2000$) flowing through a cylinder, varies periodically. According to the torque continuity equation, the angular momentum of the fluid, caused by the torque, varies periodically also, generating the Karman vortex street.

In the existing experiments of the fluid flowing through a cylinder, the external torque, $\textbf{W}_1^\textrm{i}$ , causes the circular cylinder to rotate in the fluids, shifting the direction of downwash flow. By virtue of the viscous nature of the fluid, the torque applied to the cylinder is transmitted to the fluids. Obviously, some different torques will produce a variety of vortex patterns as follows (Table 6).

1) Symmetrical vortex. When the velocity of the fluid is slow, and $Re$ is comparatively small (the viscous force is dominant), the fluid can produce the symmetrical vortexes behind a non-spinning circular cylinder. This is because the torques on both sides, produced by fluids flowing through this circular cylinder, are symmetrical. The difference of torques between two sides is close to zero. It can be considered as the result of two torques, $\textbf{W}_{1(1)}^\textrm{i}$ and $\textbf{W}_{1(2)}^\textrm{i}$ , applied simultaneously to a cylinder, with $\textbf{W}_{1(2)}^\textrm{i} = - \textbf{W}_{1(1)}^\textrm{i}$ . In other words, the amplitude and frequency of the torque, $\textbf{W}_{1(1)}^\textrm{i}$ , are equal to that of the torque, $\textbf{W}_{1(2)}^\textrm{i}$ , respectively, although their phase values are reversed.

2) Deviated vortex. When the fluid flow is slow, and the viscous force is dominant, the fluid may generate the asymmetrical vortexes behind a non-spinning circular cylinder. This is because the torques on both sides, induced by fluids flowing through this circular cylinder, are asymmetrical. The net torque of two sides is not equal to zero. The shed vortex will deviate to one side. It can be regarded as the result of two torques, $\textbf{W}_{1(1)}^\textrm{i}$ and $\textbf{W}_{1(2)}^\textrm{i}$ , applied simultaneously to a cylinder. Herein the frequency of the torque, $\textbf{W}_{1(1)}^\textrm{i}$ , is equal to that of the torque, $\textbf{W}_{1(2)}^\textrm{i}$. But the amplitudes of the two torques are different from each other, while their phase values are reversed.

3) Alternative vortex. In case we speed up the velocity of the fluid, $Re$ may increase properly a little (the viscous force is dominant still). The alternately shed vortex may be generated at the back of a spinning circular cylinder, forming the vortex street. This is because the torques on both sides, produced by fluids flowing through a spinning or asymmetrical cylinder, are non-symmetrical, so the torque is a periodic function. The pressure difference, caused by low-velocity fluids on both sides of the vortex street, will also exert a certain squeezing effect on the alternately shed vortex, to maintain the vortex pattern. It can be regarded as the result of two torques, $\textbf{W}_{1(1)}^\textrm{i}$ and $\textbf{W}_{1(2)}^\textrm{i}$ , applied simultaneously to a cylinder. Herein the amplitude and frequency of the torque, $\textbf{W}_{1(1)}^\textrm{i}$ , are equal to that of the torque, $\textbf{W}_{1(2)}^\textrm{i}$ , respectively, while the phase values of the two torques are different.

4) Biased vortex. In the aerosphere, when the top cloud flows through the asymmetric mountains, it also generates a vortex street sometimes. The torque, produced by clouds flowing through mountains, is also periodic, resulting in the angular momentum of a part of clouds to vary periodically. Similarly, when the clouds flow through some asymmetric skyscrapers, they also create a vortex street. It can be regarded as the result of two torques, $\textbf{W}_{1(1)}^\textrm{i}$ and $\textbf{W}_{1(2)}^\textrm{i}$ , applied simultaneously to a cylinder. Herein the frequency of the torque, $\textbf{W}_{1(1)}^\textrm{i}$, is equal to that of the torque, $\textbf{W}_{1(2)}^\textrm{i}$. But the amplitudes and phase values of the two torques are different, respectively.

Further, it can be found that arbitrary torque is a type of complicated case. It can be considered as the result of two torques, $\textbf{W}_{1(1)}^\textrm{i}$ and $\textbf{W}_{1(2)}^\textrm{i}$ , applied simultaneously to a cylinder. Herein the amplitude, frequency, and phase value of the torque, $\textbf{W}_{1(1)}^\textrm{i}$, are different from that of the torque, $\textbf{W}_{1(2)}^\textrm{i}$ , respectively. In a word, when the torque applied to the fluid changes, the angular momentum of the fluid element will alter, transforming the vortex street accordingly. As the elapse of time, the liquid viscosity and external torque and other factors will shift, resulting in the angular momentum of fluid elements to be constantly weakened. The angular momentum of the fluid element will gradually decrease until dissipation. Finally, the rotational fluid degenerates into an irrotational fluid.

\subsection{Zero torque}

In case the torque is equals to zero, that is, $\textbf{W}_1^\textrm{i} = 0$, the torque continuity equation will be degenerated into,
\begin{eqnarray}
\nabla \cdot ( \partial \textbf{L}_1 / \partial t ) = 0 ~,
\end{eqnarray}
one of solutions is, $\partial \textbf{L}_1 / \partial t = 0$. In other words, the angular momentum, $\textbf{L}_1$ , of fluid is a vector that does not change with time.

In the torque continuity equation, it can be considered that two operators, $\nabla$ and $ \partial / \partial t$ , are commutative. As a result, the above can be rewritten as,
\begin{eqnarray}
\partial ( \nabla \cdot \textbf{L}_1 ) / \partial t = 0 ~,
\end{eqnarray}
one of solutions is, $\nabla \cdot \textbf{L}_1 = f_b $ , with $f_b$ being a scalar. It means that the divergence of angular momentum, $\textbf{L}_1$ , of fluid is a scalar which does not change with time.

\subsection{Field strength}

According to Eq.(9), it is easy to find that the octonion field strength and octonion torque must cooperate with each other, in order to play an effective role in the fluid movements. Eq.(9) can be rewritten as,
\begin{eqnarray}
f_c + ( \partial_0 W_{10} + \nabla \cdot \textbf{W}_1^\textrm{i} ) = 0 ~,
\end{eqnarray}
where $- f_c = ( \textbf{g} \cdot \textbf{W}_1^\textrm{i} / v_0 - \textbf{b} \cdot \textbf{W}_1 ) / v_0 + k_{eg}^2 ( \textbf{E} \cdot \textbf{W}_2^\textrm{i} / v_0 - \textbf{B} \cdot \textbf{W}_2 ) / v_0$ .

a) When $f_c + \partial_0 W_{10} = 0 $ , there is $\nabla \cdot \textbf{W}_1^\textrm{i} = 0$ . Its solution is similar to that of Eq.(11). It means that there may be a few solutions for $\nabla \cdot \textbf{W}_1^\textrm{i} = 0$, including $\textbf{W}_1^\textrm{i} = 0$.

b) When $f_c + \partial_0 W_{10} \neq 0 $ , Eq.(15) has some solutions similar to that of Eq.(10). One of these solutions is, $f_c + \partial_0 W_{10} = - \partial ( \nabla \cdot \textbf{L}_1' ) / \partial t$ , with $\textbf{L}_1'$ being the equivalent angular momentum. This is one Karman vortex street under certain octonion field strengths. However the Karman vortex streets under certain octonion field strengths may be different to that without any octonion field strength.

Apparently the octonion field strength has a certain contribution to the Karman vortex streets. According to Eq.(3), the octonion field strength will contribute to the terms of octonion torque, in particular the torque $\textbf{W}_1^\textrm{i}$ and others. According to Eq.(9), the octonion field strength may directly impact the term, $( \partial_0 W_{10} + \nabla \cdot \textbf{W}_1^\textrm{i} )$, which is related to the Karman vortex streets. In other words, the Karman vortex streets can be moderately modulated and shifted by controlling the octonion field strength or octonion torque.

\section{Experiment proposal}

According to the torque continuity equation, the divergence of angular momentum and external torque are not independent of each other anymore, but closely related. On the basis of the existing experiments \cite{dynnikova,wang,sasaki} of fluid mechanics, some proposed experiments are presented in the paper, validating the torque continuity equation.

\subsection{Unidirectional torque}

In the existing experiments of fluids (water or air) flowing through a cylinder, the rotation of the cylinder with an angular velocity, $\omega_0$ , can shift the direction of downwash flow of fluids. By virtue of the viscous nature of the fluid, the torque of a rotating cylinder can be transmitted to the fluids. The torque that the fluid is subjected to is unidirectional. Similarly, the torque, produced by the airflow on the wings of an aircraft, is also the unidirectional torque approximately.

On the basis of the existing experiments, a suggested experiment relevant to the unidirectional torque can be put forward. According to the torque continuity equation, the torque that the fluid is subjected to is $\textbf{W}_1^\textrm{i}(\omega_0)$ , transforming the divergence of angular momentum of fluids. On the contrary, if the divergence of angular momentum of the fluid changes, the torque will produce correspondingly. For example, in a given streamtube with the arbitrary and varying cross section, in case the angular momentum of fluids changes with the fluid flow or time, the fluid must be affected by the external torque, $\textbf{W}_1^\textrm{i}$ . Consequently, the fluid will definitely exert a reverse torque, $-\textbf{W}_1^\textrm{i}$ , on the outer wall of the streamtube.

By means of the measurement of the reverse torque, $-\textbf{W}_1^\textrm{i}$ , on the outer wall of the streamtube, it is able to determine the influence of the external torque, $\textbf{W}_1^\textrm{i}$, on the fluids. It means that one can validate the torque continuity equation, in virtue of the measurement values of the divergence of angular momentum, $\nabla \cdot \textbf{L}_1$ , and torque $\textbf{W}_1^\textrm{i}$ .

\subsection{Reciprocating torque}

On a circular cylinder with the rotating angular velocity $\omega_1$ , there are two types of torques. One is the unidirectional torque $\textbf{W}_1^\textrm{i}(\omega_1)$ with the angular velocity $\omega_1$ , the other is the reciprocating torque, $\textbf{W}_1^\textrm{i}(\omega_2)$ , with the angular velocity $\omega_2$ . As a result, the fluids flowing through the cylinder are affected by the resultant torques, $f_{11} \textbf{W}_1^\textrm{i}(\omega_1) + f_{12} \textbf{W}_1^\textrm{i}(\omega_2)$ . For instance, what the vortex street applied to the circular cylinder is one type of reciprocating torque, and vice versa. Herein $f_{11}$ and $f_{12}$ both are coefficients.

When $\omega_1 \neq 0$ and $\omega_2 = 0$ , it is able to achieve the existing experimental results, relevant to the fluids flowing through a cylinder to produce the downwash. In case $\omega_1 = 0$ and $\omega_2 \neq 0$ , there are existing experimental results of the Karman vortex street. If $\omega_1 \neq 0$ and $\omega_2 \neq 0$ , one can obtain comparatively complicated results of the fluid flow. In general, there may be a series of angular velocities, $\omega_1$ , $\omega_2$ , $\omega_3$, ... , generating more complicated phenomena of the fluid flow than the preceding results.

The motion of the vortex street in the fluids also produces a reverse torque to exert an influence on the rotating circular cylinder. We can find the contribution of these reverse torques, by means of the measurement. Similarly, the motion of vortex street of clouds in the atmosphere yields a reverse torque effect on the mountains or skyscrapers too.

\subsection{Frequency of torque}

When $\omega_1 = 0$ and $\omega_2 \neq 0$ , it is able to alter locally the experiment results of Karman vortex street, by means of transforming the frequency, $\omega_2$ , of reciprocating torque $\textbf{W}_1^\textrm{i}(\omega_2)$ .

When the flow of the fluid is affected by the torque exerted by the outside environment, the divergence of angular momentum of the vortex street will change accordingly. The periodic variation of the torque will lead to the periodic variation of the divergence of angular momentum of the vortex street, resulting in the emergence of Karman vortex street finally. The frequency of divergence of angular momentum is equal to that of external torque. It may be $\pi / 2$ the phase difference between the external torque and the divergence of angular momentum.

If we vary the frequency $\omega_2$ of the external torque, it is able to achieve the Karman vortex street with different frequencies. Especially when $\omega_2 > \omega_0$, it is capable of acquiring the Karman vortex street with some comparatively high frequencies. As time goes on, the viscous flow will bring the damping effect, decreasing the frequency, $\omega_2$ , of the detached vortex street gradually and tending to $\omega_0$ .

These experiment proposals are helpful to further understand the physical properties of vortex street. It can be expected that scholars will utilize the physical properties of vortex street, improving the lift characteristics of various aircrafts, reducing air resistance.

\begin{table}[h]
\caption{In the octonion field theory, the brief comparisons of two torques, $\textbf{W}_{1(1)}^\textrm{i}$ and $\textbf{W}_{1(2)}^\textrm{i}$, including their phase values, frequencies, and amplitudes for some vortex patterns.}
\center
\begin{tabular}{@{}llll@{}}
\hline\hline
  vortex patterns             &   frequencies           &   amplitudes         &    phase values     \\
\hline
  symmetrical vortex          &   identical             &   identical          &    reversed         \\
  deviated vortex             &   identical             &   different          &    reversed         \\
  alternative vortex          &   identical             &   identical          &    different        \\
  biased vortex               &   identical             &   different          &    different        \\
\hline\hline
\end{tabular}\label{ta1}
\end{table}

\section{Conclusions and discussions}

In the octonion field theory, when the octonion force equals to zero, it is capable of deducing the fluid continuity equation, current continuity equation, force equilibrium equation, precession equilibrium equation, and torque continuity equation and so forth, from $\mathbb{N} = 0$. The divergence of angular momentum and external torque in the torque continuity equation are similar to the mass and linear momentum, respectively, in the fluid continuity equation.

According to the torque continuity equation, by means of transforming the magnitude and frequency of the external torque, it is able to alter that of the divergence of angular momentum of the vortex street, respectively, in the fluids. The external torque exerted on the fluids is related to the velocity circulation. Obviously, remodeling the motion mode of vortex street can improve the lift and drag characteristics produced by the fluids.

In the octonion field theory, the preceding continuity equations and equilibrium equations will be affected by the electromagnetic strength and gravitational strength. Further, in case each of electromagnetic strength and gravitational strength is equal to zero, the fluid continuity equation and current continuity equation, in the octonion field theory, will be simplified into their conventional equations respectively, in the classic fluid mechanics.

In the octonion space, when there is the relative motion between two inertial coordinate systems, not only the conservation law of mass but also conservation law of charge can be derived from the invariant under the coordinate transformations. By comparison, one can find some characteristics as follows. a) The conservation law of mass and conservation law of charge are independent of the fluid continuity equation and current continuity equation, respectively. b) The precondition for the establishment of the conservation law of mass is different from that of the fluid continuity equation. Although both of them can be established simultaneously, there is no logical necessary connection between them. c) Similarly, the precondition for the establishment of the conservation law of charge is different from that of the current continuity equation. Either there is no logical necessary connection between both of them, although the two can be established simultaneously (in Table 4).

However, in the classic field theory, the conventional fluid continuity equation can be derived from the conservation law of mass, from a comparatively simple point of view. Unfortunately, it is unable to consider the influence of electromagnetic strength and gravitational strength on the conventional fluid continuity equation. This inference obviously constitutes a major defect of the conventional fluid continuity equation in the classic fluid mechanics, limiting the scope of application of the conventional fluid continuity equation. Similarly, the conventional current continuity equation can be derived from the conservation law of charge, from a comparatively simple point of view. Its defects confine the application scope of the conventional current continuity equation in the classic electromagnetic theory. What is more serious is that classic field theory is incapable of discovering some equations, such as the precession equilibrium equation and torque continuity equation and so forth in the octonion field theory (in Table 5).

In the classic fluid mechanics, the fluid continuity equation, linear momentum equation, and energy equation constitute the significant theoretical cornerstone. The torque continuity equation studied in this paper is able to describe some new characteristics of the fluid motion, including the physical properties of the torque and angular momentum of fluid elements. In the future, it is necessary to consider the torque continuity equation, besides the fluid continuity equation, linear momentum equation, and energy equation and others, when we study the phenomena of fluid flow. It can be expected that the torque continuity equation will promote the follow-up development of the fluid mechanics to some extent.

It is worth noting that the paper discusses only some effects of the external torque and divergence of angular momentum of fluids on the torque continuity equation. The torque applied to the fluids can change the angular momentum and velocity circulation of the fluids. The divergence of angular momentum and torque comply with the requirements of torque continuity equation. This is as if the mass and linear momentum obey the fluid continuity equation. In the octonion field theory, the motion mode of fluids can simultaneously satisfy both the torque continuity equation and the fluid continuity equation in general. In the future research, we shall explore the contribution of electromagnetic strength and gravitational strength to the torque continuity equation in the plasma and magnetofluid. And it is going to carry out the influence of the external torque with time-varying frequency on the motion of fluid flow. This will help to deepen the understanding of the motion properties of the vortex street of fluids, improving the lift performances and drag characteristics of aircrafts, promoting the further development of the fluid mechanics.

\section*{Acknowledgments}
The author is indebted to the anonymous referees for their valuable comments on the previous manuscripts. This project was supported partially by the National Natural Science Foundation of China under grant number 60677039.

\subsection*{Conflict of interest}
The author declares no potential conflict of interests.

\end{document}